\documentclass[fleqn,usenatbib, usenames,dvipsnames]{mnras}
\usepackage[T1]{fontenc}
\DeclareRobustCommand{\VAN}[3]{#2}
\let\VANthebibliography\thebibliography
\def\thebibliography{\DeclareRobustCommand{\VAN}[3]{##3}\VANthebibliography}
\usepackage{graphicx}	
\usepackage{amsmath}	
\usepackage{amssymb}	
\usepackage{xcolor}
\usepackage{color}
\usepackage{soul}
\usepackage{adjustbox} 
\newcommand{\lt}{\ensuremath <}

\newcommand{\change}[1]{\textcolor{black}{#1}}
\title[Radio imaging of lensed quasars]{Radio imaging of gravitationally lensed radio-quiet quasars}
\author[Jackson et al.]{
Neal Jackson$^{1}$\thanks{E-mail: neal.jackson@manchester.ac.uk},
Shruti Badole$^{1}$,
Thomas Dugdale$^{1}$,
Hannah R. Stacey$^{2}$,
Philippa Hartley$^{3}$\\
{\rm \Large and J. P. McKean$^{4,5,6}$} \\
$^{1}$Jodrell Bank Centre for Astrophysics, Department of Physics and Astronomy, University of Manchester, Oxford Rd, Manchester M13 9PL, UK\\
$^{2}$ European Southern Observatory, Karl-Schwarzschild Str. 2, D-85748 Garching bei M\"unchen, Germany\\
$^{3}$ SKA Observatory, Jodrell Bank, Macclesfield SK11 9FT, UK \\
$^{4}$Kapteyn Astronomical Institute, University of Groningen, Postbus 800, NL-9700 AV Groningen, The Netherlands\\
$^{5}$South African Radio Astronomy Observatory (SARAO), P.O. Box 443, Krugersdorp 1740, South Africa\\
$^{6}$Department of Physics, University of Pretoria, Lynnwood Road, Hatfield, Pretoria, 0083, South Africa\\
}
\date{Accepted XXX. Received YYY; in original form ZZZ}
\pubyear{2015}
\begin{document}
\maketitle
\begin{abstract}
We present 6-GHz Very Large Array radio images of 70  gravitational lens systems at 300-mas resolution, in which the source is an optically-selected quasar, and nearly all of which have two lensed images. We find that about in half of the systems (40/70, with 33/70 secure), one or more lensed images are detected down to our detection limit of 20\,$\mu$Jy\,beam$^{-1}$, similar to previous investigations and reinforcing the conclusion that typical optically-selected quasars have intrinsic GHz radio flux densities of a few $\mu$Jy ($\sim10^{23}$\,W\,Hz$^{-1}$ at redshifts of 1--2). In addition, for ten cases it is likely that the lensing galaxies are detected in the radio. Available detections of, and limits on the far-infrared luminosities from the literature, suggest that nearly all of the sample lie on the radio-FIR correlation typical of star-forming galaxies, and that their radio luminosities are at least compatible with the radio emission being produced by star formation processes. One object, WISE2329$-$1258, has an extra radio component that is not present in optical images, and is difficult to explain using simple lens models. In-band spectral indices, where these can be determined, are generally moderately steep and consistent with synchrotron processes either from star-formation/supernovae or AGN. Comparison of the A/B image flux ratios at radio and optical wavelengths suggests a 10 per cent level contribution from finite source effects or optical extinction to the optical flux ratios, together with sporadic larger discrepancies that are likely to be due to optical microlensing.
\end{abstract}
\section{Introduction}

It has been clear for some decades that the influence of active galactic nuclei (AGN), and their accompanying central super-massive black holes, is important for the evolution of galaxies. Galaxies are thought to form when gas collapses within dark matter haloes (e.g. \citealt{1978MNRAS.183..341W}). Simple models predict that significant star formation should \change{occur} within halos of a wide range of masses. In particular, these models overpredict the star formation rate in high-mass galaxies, leading to much more massive and luminous galaxies than are actually observed. This problem can be solved by a range of mechanisms collectively known as ``feedback''. At high halo masses, this feedback consists of the influence of an AGN, which injects energy and momentum into the interstellar medium and thereby suppresses star formation \citep{1998A&A...331L...1S,2005MNRAS.361..776S,2006MNRAS.365...11C}. The details of how this feedback operates are relatively complicated; it may proceed either at high or low rates of accretion of the central black hole, and the duty cycle (the fraction of time during which the feedback is operating) can also vary according to the mode of accretion \citep{2005MNRAS.362...25B,2012MNRAS.421.1569B}. Evidence for the feedback model includes a tight correlation between black-hole mass and properties of the wider galaxy such as stellar velocity dispersions \citep{2000ApJ...539L...9F} and more detailed studies of individual objects (e.g. \citealt{2010A&A...521A..65N,Rupke:2017,Murthy:2022,Girdhar:2022}).

Therefore, it is important to understand the properties of galaxies containing AGN, particularly those containing quasars, the most energetic AGN.  High-resolution studies at radio wavelengths can make an important contribution to this effort. Firstly, in a minority of quasars, there is strong radio emission, providing direct evidence for relativistic jets from the AGN that may remove gas from within the stellar bulge (e.g. \citealt{Girdhar:2022}). Secondly, high-resolution observations can definitively prove the presence of an AGN if components with brightness temperatures greater than about $10^5$\,K are found \citep{1990ApJ...359..291N,1992ARA&A..30..575C,2022MNRAS.515.5758M}; or, alternatively, suggest the dominance of star-formation if the distributed radio emission is coincident with dust, as indicated by rest-frame Far-Infrared (FIR) continuum emission (e.g. \citealt{2020MNRAS.496..138B}).

It is not clear whether radio-loud and radio-quiet quasars are separate populations\footnote{See \cite{2016A&ARv..24...13P} for an argument that ``radio-quiet'' as a designation should be replaced by ``unjetted'' in the sense of strong relativistic jets, emitting strong radio emission and $\gamma$-rays, not being present.}. Initial claims of a dichotomy in radio luminosity \citep{1989AJ.....98.1195K} have been variously supported \citep{1990MNRAS.244..207M,2007ApJ...656..680J} and questioned \citep{2003MNRAS.346..447C,2013ApJ...764...43S} with some quasars being radio-silent to a very high degree \citep{2021A&A...649L...9R}. 

Whatever the truth, it is likely that radio emission both from AGN synchrotron emission and from supernova/HII regions associated with star-formation processes are at least partly present in radio-weak AGN. Evidence for star-formation processes includes the form of the radio flux density distributions for optically-selected quasars \citep{2013ApJ...768...37C}, and the positive correlation of star-formation rates, inferred from FIR data, with radio luminosities in a faint radio sample \citep{2015MNRAS.453.1079B}. This latter evidence is an extension of the observation that radio and FIR luminosities in star-forming objects correlate extremely well over a wide range in luminosity \citep{1991MNRAS.251P..14S}, with radio-excesses above this correlation being expected only in objects with a significant AGN contribution.

On the other hand, excess radio emission above that expected from star-forming processes is observed in a faint FIR-selected survey \citep{2017MNRAS.468..217W}, suggesting an AGN contribution; and high-resolution imaging has given significant numbers of detections of compact radio structure in faint radio sources from optically-selected surveys \citep{2017A&A...607A.132H,Radcliffe:2018}. More recently, \cite{2023MNRAS.518...39W} \change{inferred that both emission mechanisms operate at some level} in low-redshift Palomar-Green survey quasars. Investigation of such objects \citep{2022ApJ...936...73A,2023arXiv230713599C} reveals the frequent presence of milliarsecond-scale cores of brightness temperature $\sim 10^7$\,K \citep{2023arXiv230713599C}, with some evidence for emission from the corona above the accretion disk rather than explicitly from an AGN radio jet. \change{Using LOFAR DR1 data, \cite{2023arXiv231210177C} find a detection rate in the radio of up to 94\% in a set of optically-selected quasars, using the deepest LOFAR data, and derive an AGN excess in the majority of objects by comparing LOFAR and far-infrared fluxes.}

High-resolution radio imaging offers, in principle, a clean test of emission mechanisms. This is generally very difficult due to the extreme radio faintness of high-redshift radio-quiet quasars. However, considerable work has now been done using gravitationally lensed quasars. \change{Here, typical lensing magnification of factor 5--10, together with the linear scaling of flux density with magnification factor, allows us to reach objects with an order of magnitude lower intrinsic radio flux density levels} with relative ease \citep{2011ApJ...739L..28J,2015MNRAS.454..287J,2019MNRAS.485.3009H,2019A&A...622A..18S,2020MNRAS.496..138B,2021MNRAS.505L..36M,2021MNRAS.508.4625H,2021MNRAS.508L..64M}. The results are mixed, with clear evidence for high-brightness temperature radio components in some cases \citep{2019MNRAS.485.3009H} and some objects having radio emission ascribable to star formation \citep{2020MNRAS.496..138B}. \change{A summary of existing information, with some new data, is given by \cite{2021MNRAS.508.4625H}}.

Radio flux densities are now available for many lensed radio-quiet quasars, which is an essential preliminary for follow-up studies with higher-resolution telescopes. Radio imaging also allows a first-look comparison of radio and FIR luminosities to make an initial assessment of the likelihood of AGN/non-AGN origins for the radio emission. The most recent such study \citep{2023arXiv231107836D} gives detections of about 50 percent of a sample of 24 radio-quiet Gaia Gravitational Lenses (GraL) quasars at levels of a few tens of $\mu$Jy\,beam$^{-1}$, typical of other studies at similar flux density levels which yield detections at a few tens of percent. Many observations to date have concentrated on four-image lenses with an optically selected quasar as the source. Here, we present Karl G. Jansky Very Large Array (VLA) data for a sample of predominantly two-image lens systems with optically-selected quasars. We aim to detect radio emission from as many sources as possible, to make a preliminary determination of any objects whose radio flux density exceeds that expected from purely star-forming processes and, where possible, to gain further information on the emission mechanisms using spectral indices and comparison of the radio and optical flux ratios. Further Very Long Baseline Interferometry (VLBI) follow-up can then be obtained as necessary for those objects in which the presence of significant levels of radio emission by AGN is suspected (e.g. \cite{2019MNRAS.485.3009H}). In Section~\ref{section:observations} we describe the observations and present the radio images and detection statistics for the lensed quasar sample, with descriptions of interesting individual objects, and in Section~\ref{section:discussion} we discuss the possible physical mechanisms for the radio emission by comparison with other wavebands. Throughout, we assume a flat $\Lambda$CDM Universe with $H_0=67.4\,$km\,s$^{-1}$Mpc$^{-1}$ with $\Omega_{\rm m}$=0.31 \citep{2020A&A...641A...6P}. We define $S_\nu\propto\nu^{\alpha}$ for the relation between flux density $S_\nu$ at frequency $\nu$ and spectral index $\alpha$.

\section{Observations and results}
\label{section:observations}
\subsection{Sample selection and observations}

The sample for observation was chosen from the list of known lensed radio quasars maintained by Lemon et al.\footnote{\tt https://research.ast.cam.ac.uk/lensedquasars}. Known radio-loud gravitational lens systems, mainly from the CLASS, PMN and MG surveys \citep{2003MNRAS.341...13B,2000AJ....120.2868W,1992AJ....104..968H} were excluded, as were 4-image systems, many of which have already been observed in the radio \citep{2011ApJ...739L..28J,2015MNRAS.454..287J,2019MNRAS.485.3009H,2020MNRAS.496..138B,2021MNRAS.508.4625H}. The observations were conducted in two observing cycles (2020 and 2023) during the period of observations scheduled for A-configuration in each case. In the 2020 period, objects were selected with declinations between $-20^{\circ}$ and $+25^{\circ}$, to permit observations with both the VLA and ALMA. 62 objects resulted from this process, of which 45 were observed based on the availability of observing time at different LST. In the second period, objects at all declinations $>-30^{\circ}$ were selected and 25 further objects were observed according to their availability as a function of LST.

Each observation was conducted for a total of 22.5 min on source, with scans of 4.5 min interspersed with 1.5-min observations of a nearby, bright phase calibration source to correct for the time-varying phase contribution due to the atmosphere above each antenna. This on-source time gives a theoretical r.m.s. noise level of about 5\,$\mu$Jy\,beam$^{-1}$ with natural weighting. Standard flux calibrator sources (3C48, 3C138 and 3C286) were observed immediately before or after each source observation to provide absolute flux calibration and calibration of the bandpass response. Most objects were observed in pairs using a common flux calibrator, and, where possible, a common phase calibrator. Observations were conducted using a 4-GHz bandwidth between frequencies of 4 and 8~GHz. 2-second integration times and 1-MHz channel widths were used, although these were averaged during further analysis as only the central few arcseconds of each field of view were of scientific interest.

\begin{figure}
    \begin{tabular}{cc}
    \includegraphics[width=0.44\columnwidth]{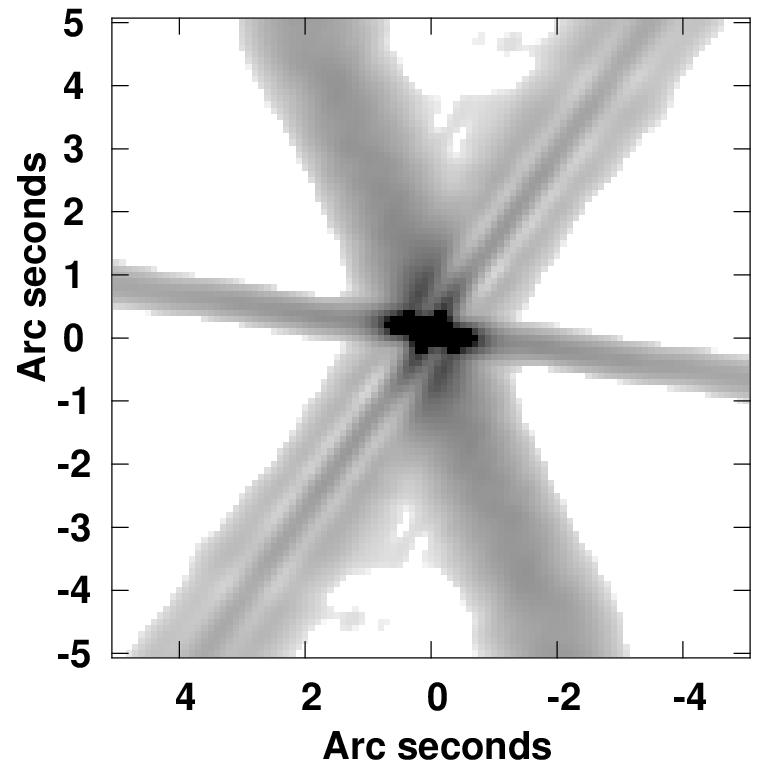}&
    \includegraphics[width=0.46\columnwidth]{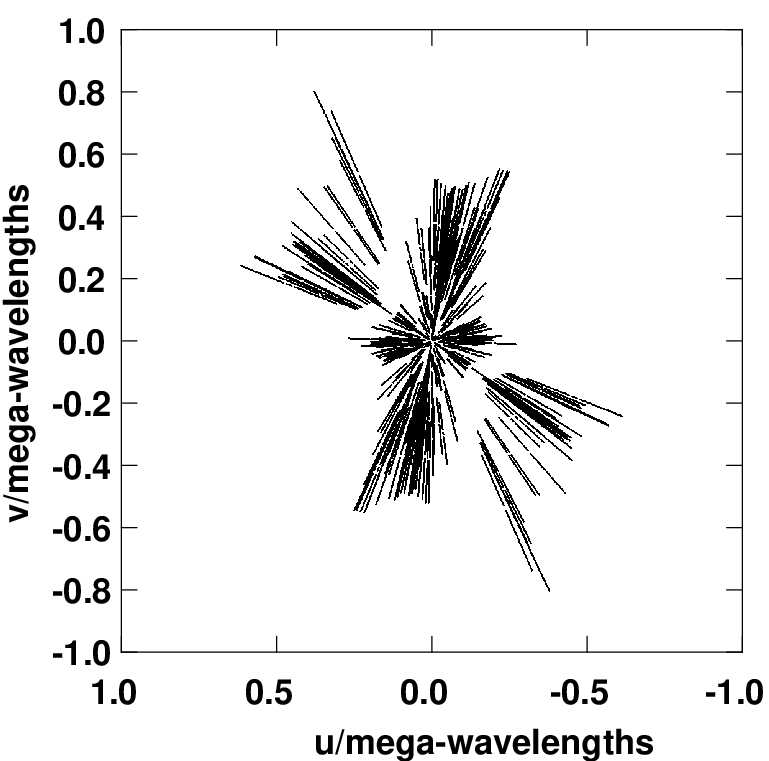}\\
    \end{tabular}     
    \caption{Right: $uv$ plane coverage for a typical snapshot observation with the hybrid A/AnB configuration used during the 2020 observations, at declination +20$^{\circ}$N. Left: point-spread function (dirty beam) resulting from this $uv$ coverage. }
     \label{uvandbeam}
\end{figure}

The first group of observations were conducted between 2020 November 19 and 2020 December 1. For operational reasons due to the COVID pandemic, the VLA was, during this time, in a non-standard configuration resembling the AnB intermediate configuration. This configuration consisted of the north arm fully extended in A-configuration, the east arm in the more compact B-configuration, and the west arm in the B-configuration except for two antennas that had been moved to the end positions (W64 and W72) of the west arm. Fig. \ref{uvandbeam} shows the $uv$ coverage, together with the point spread function, for a typical target observation with a source at declination $+20^{\circ}$. 
The effects of the non-uniform distribution of antennas in each of the three arms of the interferometer are clearly visible and lead to a generally higher level of beam artefacts in each of the maps. In 2023, observations were conducted using the standard A-configuration. The typical beam size of the images, with natural weighting, is about 0\farcs5 at high declinations and larger (0\farcs7--1\farcs0) at lower declinations. A $z=1$ source with a flux density of 100\,$\mu$Jy and an intrinsic size equivalent to the beam size has, at these observing frequencies, a brightness temperature of 40\,K, so lower limits on brightness temperature derived from these observations on their own do not give constraints on the origin of radio emission.

\subsection{Data reduction}

Observations were flagged using the automatic flagger in the {\sc casa} package \citep{2022PASP..134k4501C}, distributed by the U.S. National Radio Astronomy Observatory (NRAO), which applies flags based on auto-calculated thresholds using RMS values in regions of time and frequency. A relatively low \change{flagging threshold ($\sigma=3$) was used to remove radio frequency interference}. Data were then read into the {\sc aips} package of the NRAO \citep{2003ASSL..285..109G} and processed using {\sc parseltongue} scripts \citep{2006ASPC..351..497K}. In \change{several} cases, it was found that the {\sc casa} auto-flagging severity, which was required to remove bad data, had also removed the flux calibrator scan, in cases where the telescopes were not on source for some of the observation and the visibilities changed from zero to a high value. The analysis pipeline was adjusted to reinstate the on-source parts of these scans. Data were then averaged in time and frequency by a factor of 4, and calibrated using fringe-fitting to remove delays. Initial bandpass calibration was done using the flux calibrator scan, and then amplitude and phase calibration was performed using the phase calibrator together with the flux calibrator to adjust the flux scale. All calibration was then copied back to the unaveraged data, which was used for the imaging to reduce the effects of bandwidth and integration time smearing on other sources in the field. Imaging of the calibrated data was performed in {\sc casa} using multi-frequency synthesis (MFS) and natural weighting of the data; this weighting gives the maximum signal-to-noise ratio in the final images at the expense of resolution and shape of the beam.  Most observations were affected by nearby bright radio sources, some severely so. Hence, nearby bright sources were identified by use of the FIRST survey \citep{1995ApJ...450..559B}, for objects within the FIRST footprint, and imaged in a multi-field deconvolution together with the area around the target source. One source outside the FIRST footprint, PS\,J2332$-$1852, has been deconvolved taking into account bright sources from the NVSS survey \citep{1998AJ....115.1693C}. In one observation (containing SDSS\,J1515+1511 and ULAS\,J1529+1038) no flux calibrator observation is available; in this case, bootstrapping was done from published flux densities of the phase calibrators, but examination of the noise level in the images suggests that the flux scale is approximately 50\% too high for these objects. {\sc casa} images were compared with those made in {\sc aips} without MFS; nearly all are very similar in both the visual appearance of the images and derived flux densities, except for SDSS\,J1258+1657 where the {\sc aips} image was used. Images were also made with {\sc robust}=0 weighting  \citep{1995AAS...18711202B} but these generally gave much poorer detection rates compared to naturally-weighted images. Final naturally-weighted images are shown in Fig.~\ref{images}, together with optical contours from the Pan-STARRS public data \citep{2020ApJS..251....7F}. Point images detected by {\it Gaia} \citep{2018A&A...616A...1G} are also overplotted on the radio images; since both the radio and {\it Gaia} astrometric frames are more accurate than that of Pan-STARRS, the Pan-STARRS images have been shifted by eye to correspond with the {\it Gaia} frame. The VLA astrometry should be good to about 10-20~mas, similar to that of the phase calibrator network \citep{1992MNRAS.254..655P}, with the {\it Gaia} astrometry much better than this.




\begin{table*}
\caption{Observed objects with the source and lens redshifts (where known). Taken from the compilation of C. Lemon ({\tt https://research.ast.ac.ac.uk/quasars}, and references therein). References to original discovery papers: A18 = \protect\cite{2018MNRAS.479.4345A}; AA18 = \protect\cite{2018MNRAS.475.2086A}; AN18 = \protect \cite{2018MNRAS.480.5017A}; C01 =\protect \cite{2001AJ....122.1679C}; H99 = \protect\cite{1999A&AS..134..483H}; H00 = \protect\cite{2000A&A...357L..29H}; I05 = \protect\cite{2005AJ....130.1967I}; I06 = \protect\cite{2006AJ....131.1934I}; I07 = \protect\cite{2007AJ....133..206I}; I08 = \protect\cite{2008AJ....135..496I}; I09 = \protect\cite{2009AJ....137.4118I}; I14 = \protect\cite{2014AJ....147..153I}; J08 = \protect\cite{2008MNRAS.387..741J}; J09 = \protect\cite{2009MNRAS.398.1423J}; J12 = \protect\cite{2012MNRAS.419.2014J}; K10 = \protect\cite{2010AJ....139.1614K}; K19 = \protect\cite{2019arXiv191208977K}; L18 = \protect\cite{2018MNRAS.479.5060L}; L19 = \protect\cite{2019MNRAS.483.4242L}; M10 = \protect\cite{2010AJ....140..370M}; M16 = \protect\cite{2016MNRAS.456.1595M}; O08 = \protect\cite{2008AJ....135..520O}; P03 = \protect\cite{2003AJ....125.2325P}; R11 = \protect \cite{2011ApJ...738...30R}; R13 = \protect\cite{2013ApJ...765..139R}; S17 = \protect\cite{2017AJ....153..219S}; W18 = \protect\cite{2018MNRAS.477L..70W}; W96 = \protect \cite{1996A&A...315L.405W}; W99 = \protect \cite{1999A&A...348L..41W}.}
\vspace{-8pt}
\setlength{\tabcolsep}{5pt}
\change{
\begin{tabular}{lccc p{5pt} lccc p{5pt} lccc }
    \cline{0-3} \cline{6-9} \cline{11-14} \rule{-1.5pt}{2.2ex}   
    Object & $z_s$ & $z_l$ & Refs & &Object & $z_s$ & $z_l$ & Refs & & Object & $z_s$ & $z_l$ & Refs\\ 
    \cline{0-3} \cline{6-9} \cline{11-14} \rule{-1.5pt}{3ex}   
    J0011$-$0845 & 1.7 & - & L18 & & ULASJ0820+0812 & 2.02 & 0.80 & J09 & & SDSSJ1458$-$0202 & 1.72 & - & M16\\
    J0013+5119 & 2.63 & - & L19 & & SDSSJ0832+0404 & 1.12 & 0.66 & O08 & & SDSSJ1515+1511 & 2.06 & 0.74 & I14\\
    PSJ0028+0631 & 1.06 & - & L18 & & J0907+0003 & 1.30 & - & K19 & & ULASJ1527+0141 & 1.44 & 0.30 & J12\\
    J0102+2445 & 2.09 & 0.27 & L19 & & J0941+0518 & 1.54 & 0.34 & W18,L18 & & ULASJ1529+1038 & 1.97 & 0.40 & J12\\
    SDSSJ0114+0722 & 1.83 & 0.41 & M16 & & SDSSJ0946+1835 & 4.80 & 0.39 & M10 & & J1616+1415 & 2.88 & - & L19\\
    J0124$-$0033 & 2.84 & - & L19 & & SDSSJ1029+2623 & 2.20 & 0.58 & I06 & & SDSSJ1620+1203 & 1.16 & 0.40 & K10\\
    J0140$-$1152 & 1.80 & 0.28 & L18 & & SDSSJ1054+2733 & 1.45 & 0.23 & K10 & & J1623+7533 & 2.64 & - & L19\\
    PSJ0140+4107 & 2.50 & - & L18 & & SDSSJ1055+4628 & 1.25 & 0.39 & K10 & & PSJ1640+1045 & 1.70 & - & L18\\
    J0146$-$1133 & 1.44 & - & L18,A18 & & SDSSJ1128+2402 & 1.61 & - & I14 & & PSJ1831+5447 & 1.07 & - & L18\\
    J0203+1612 & 2.18 & - & L19 & & SDSSJ1131+1915 & 2.92 & 0.32 & K10 & & J1949+7732 & 1.63 & - & L19\\
    J0228+3953 & 2.07 & - & L19 & & SDSSJ1226$-$0006 & 1.12 & 0.52 & I02,P03 & & PSJ2124+1632 & 1.28 & - & L18\\
    J0235$-$2433 & 1.44 & - & L18 & & SDSSJ1254+1857 & 1.72 & 0.56 & I09 & & HE2149$-$2745 & 2.03 & 0.60 & W96\\
    HE0230$-$2130 & 2.16 & 0.52 & W99 & & SDSSJ1254+2235 & 3.63 & 0.30 & M16 & & A2213$-$2652 & 1.27 & - & AA18\\
    DESJ0245$-$0556 & 1.54 & - & A18 & & SDSSJ1258+1657 & 2.70 & 0.40 & I09 & & HS2209+1914 & 1.07 & - & H99\\
    J0246$-$1845 & 1.86 & - & K19,L19 & & SDSSJ1304+2001 & 2.18 & 0.40 & K10 & & J2250+2117 & 1.73 & - & L19\\
    SDSSJ0246$-$0825 & 1.69 & 0.72 & I05 & & SDSSJ1320+1644 & 1.50 & 0.90 & R13 & & J2257+2349 & 2.11 & - & W18\\
    SDSSJ0256+0153 & 2.60 & 0.61 & M16 & & SDSSJ1334+3315 & 2.43 & 0.56 & R11 & & PSJ2305+3714 & 1.78 & - & L18\\
    DESJ0407$-$1931 & 1.52 & - & AN18 & & SDSSJ1339+1310 & 2.24 & 0.61 & I09 & & PSS2322+1944 & 4.12 & 1.23 & C01\\
    SDSSJ0737+4825 & 2.89 & 1.54 & M16 & & SDSSJ1349+1227 & 1.72 & 0.65 & K10 & & WISE2329$-$1258 & 1.31 & 1.15 & S17\\
    ULASJ0743+2457 & 2.17 & 0.38 & J12,I14 & & SDSSJ1353+1138 & 1.62 & 0.25 & I06 & & PSJ2332$-$1852 & 1.49 & - & L18\\
    SDSSJ0746+4403 & 2.00 & 0.51 & I07 & & ULASJ1405+0959 & 1.81 & 0.66 & J12 & & ULASJ2343$-$0050 & 0.79 & 0.30 & J08\\
    SDSSJ0806+2006 & 1.54 & 0.57 & I06 & & SDSSJ1442+4055 & 2.58 & 0.28 & M16,S16 & & J2350+3654 & 2.09 & - & L19\\
    SDSSJ0818+0601 & 2.35 & 1.01 & M16 & & SDSSJ1452+4224 & 4.82 & 0.38 & M16 & &   &   &   &  \\
    HS0818+1227 & 3.11 & 0.39 & H00 & & SDSSJ1455+1447 & 1.42 & 0.42 & K10 & &   &   &   &  \\
 \end{tabular} 

} 
\label{table:ztable}
\end{table*}

Table \ref{obstable} gives details of the observed sources, together with their observed 6-GHz flux density. This was measured using a Gaussian fit to points in the image identified by eye, allowing the peak and positions to vary while keeping the width fixed to the point spread function using the {\sc jmfit} function in {\sc aips}. Without this the fit routinely becomes unstable in cases of low signal-to-noise ratio, \change{but a few objects (WISE\,2329$-$1258, HS\,2209+1914, J2250+2117) were fitted by hand with extended sources after examination of the residuals from the automatic fitter.} In most cases, the observed off-source noise level in the images was between 4 and 7\,$\mu$Jy\,beam$^{-1}$. Images were examined by eye, with probable detections of at least one radio component from the lens system in 40/70 observations. However, a number of these are very marginal detections. To quantify the significance of these marginal detections, random positions in a typical image were fitted using the same procedure as used for the identified ``detections''. In 10\% of such cases, point-source flux densities of $>20\,\mu$Jy were returned by the fitting routine, with this proportion falling to $<$1\% at $60\,\mu$Jy.  In seven cases (J0203+1612, SDSS\,J0256+0153, SDSS\,J0806+2006, SDSS\,1254+1857, SDSS\,J1304+2001 SDSS\,J1620+1203 and J2250+2117) the only radio  detection is, or is likely to be, of radio emission from the lensing galaxy rather than the lensed quasar. In all other cases, secure detections are obtained of lensed images of the background quasar. 

\begin{table*}
\caption{The lensed quasar systems observed in this work, with coordinates of the centres of the images in Fig.~\ref{images}, VLA flux density measurements for each component at 6\,GHz from these observations in cases where radio components are detected (see text). In the case of multiple components, flux densities are given \change{in the columns marked cpt1, cpt2 and cpt3 and labelled by their orientation in the map}.  Errors on total flux densities are subject to an additional systematic uncertainty on flux calibration of about 10\%, in addition to the formal errors given in the table from the output of the modelling. Far-infrared luminosities and their origins are discussed in Section~\ref{section:discussion}; quantities in brackets are the upper and lower 1-$\sigma$ errors in the last two digits of the value. In the comments column, LG indicates that the radio emission originates in the lensing galaxy or galaxies, and M indicates a marginal/tentative detection (at least one component at or below 20\,$\mu$Jy.)}
\label{obstable}
\begin{tabular}{lcccccl}\hline
Source name & Coordinates & Flux density & Flux density &Flux density & $\log_{\rm 10}(L_{\rm FIR}/L_{\odot})$ & Comments \\
&(RA, Dec J2000)&(cpt 1,$\mu$Jy)&(cpt 2,$\mu$Jy)&(cpt 3,$\mu$Jy)&&\\ \hline
  J0011$-$0845 & 00:11:20.24 $-$08:45:51.5  &     23$\pm$5  & -  & - & - & M \\
    J0013+5119 & 00:13:23.54 +51:19:05.9  &    192$\pm$4 N &    261$\pm$4 S &    302$\pm$4 C& - & 2im+LG \\
  PSJ0028+0631 & 00:28:22.49 +06:31:54.1 &      &      &     & - &   \\
    J0102+2445 & 01:02:47.22 +24:45:15.8  &     19$\pm$4  & -  & - & - & M \\
SDSSJ0114+0722 & 01:14:38.35 +07:22:28.9 &      &      &     & - &   \\
  J0124$-$0033 & 01:24:57.46 $-$00:33:11.5 &      &      &     & - &   \\
  J0140$-$1152 & 01:40:03.00 $-$11:52:19.4  &     20$\pm$5  & -  & - & - & M \\
  PSJ0140+4107 & 01:40:49.01 +41:07:59.9  &     14$\pm$4  NE&     24$\pm$4 SW & - & - & M \\
  J0146$-$1133 & 01:46:32.86 $-$11:33:38.9  &     32$\pm$5 NE &     26$\pm$5 SW & - & - &   \\
    J0203+1612 & 02:03:59.40 +16:12:08.6  &     46$\pm$4  & -  & - & - & LG \\
    J0228+3953 & 02:28:11.10 +39:53:07.3  &     42$\pm$5 E &     71$\pm$5 W & - & - &   \\
 HE0230$-$2130 & 02:32:33.19 $-$21:17:25.7  &     30$\pm$5 N &     48$\pm$5 C &     30$\pm$5 SW& 13.1(1,1) &   \\
  J0235$-$2433 & 02:35:27.41 $-$24:33:13.7  &     26$\pm$6 N &     32$\pm$6 S & - & - &   \\
DESJ0245$-$0556 & 02:45:25.56 $-$05:56:59.6  &     42$\pm$5 NE &     27$\pm$5 SW & - & - &   \\
  J0246$-$1845 & 02:46:12.20 $-$18:45:05.0  &     80$\pm$7 N &     45$\pm$7 S & - & - &   \\
SDSSJ0246$-$0825 & 02:46:34.09 $-$08:25:36.1 &      &      &     & 12.8(1,1) &   \\
SDSSJ0256+0153 & 02:56:40.73 +01:53:29.4  &    114$\pm$5  & -  & - & - & LG;$\alpha=-0.39\pm0.20$ \\
DESJ0407$-$1931 & 04:07:53.79 $-$19:31:22.1 &      &      &     & - &   \\
SDSSJ0737+4825 & 07:37:08.71 +48:25:51.2 &      &      &     & - &   \\
ULASJ0743+2457 & 07:43:52.61 +24:57:43.6  &     22$\pm$4  & -  & - & - & M \\
SDSSJ0746+4403 & 07:46:53.04 +44:03:51.3 &      &      &     & $<$12.50 &   \\
SDSSJ0806+2006 & 08:06:23.68 +20:06:31.5  &     27$\pm$5  & -  & - & 12.4(4,2) & LG \\
SDSSJ0818+0601 & 08:18:30.43 +06:01:37.6  &    161$\pm$5 NE &     48$\pm$5 SW & - & - & $\alpha=-1.02\pm0.39$ \\
   HS0818+1227 & 08:21:38.90 +12:17:30.9 &      &      &     & 12.6(4,4) &   \\
ULASJ0820+0812 & 08:20:16.09 +08:12:16.8  &    107$\pm$4  & -  & - & 12.60(04,05) &   \\
SDSSJ0832+0404 & 08:32:17.06 +04:04:04.4 &      &      &     & $<$12.10 &   \\
    J0907+0003 & 09:07:10.49 +00:03:21.2 &      &      &     & - &   \\
    J0941+0518 & 09:41:22.54 +05:18:23.8  &    113$\pm$4 SE &     70$\pm$4 NW & - & - & $\alpha=-0.11\pm0.16$ \\
SDSSJ0946+1835 & 09:46:04.84 +18:35:40.3  &     33$\pm$4  & -  & - & - &   \\
SDSSJ1029+2623 & 10:29:13.84 +26:23:30.2 &      &      &     & 12.70(1,1) &   \\
SDSSJ1054+2733 & 10:54:40.89 +27:33:06.1  &     32$\pm$5 E &    100$\pm$5 W & - & 12.40(1,1) &   \\
SDSSJ1055+4628 & 10:55:45.45 +46:28:39.7 &      &      &     & $<$12.10 &   \\
SDSSJ1128+2402 & 11:28:18.49 +24:02:17.4  &     51$\pm$4 NE &     85$\pm$4 SW & - & - & $\alpha=-0.72\pm0.20$ \\
SDSSJ1131+1915 & 11:31:57.79 +19:15:27.4  &     35$\pm$4  & -  & - & $<$12.70 &   \\
SDSSJ1226$-$0006 & 12:26:08.02 $-$00:06:02.2  &     31$\pm$4 E &     32$\pm$4 W & - & - &   \\
SDSSJ1254+1857 & 12:54:40.34 +18:57:12.0  &    226$\pm$18  & -  & - & - & LG? \\
SDSSJ1254+2235 & 12:54:19.00 +22:35:36.0 &      &      &     & - &   \\
SDSSJ1258+1657 & 12:58:19.24 +16:57:17.6  &     52$\pm$4 E &     69$\pm$4 W & - & 12.9(1,1) &   \\
SDSSJ1304+2001 & 13:04:43.60 +20:01:05.0  &     25$\pm$4  S&     16$\pm$4 N & - & 12.4(2,1) & LG \\
SDSSJ1320+1644 & 13:20:59.47 +16:44:03.7  &    123$\pm$5 E &     22$\pm$5 W & - & - & $\alpha=-0.80\pm0.18$ \\
SDSSJ1334+3315 & 13:34:01.39 +33:15:34.3 &      &      &     & - &   \\
SDSSJ1339+1310 & 13:39:07.20 +13:10:39.0  &     25$\pm$5 SE &     20$\pm$5 NW & - & 12.60(3,2) & M \\
SDSSJ1349+1227 & 13:49:29.93 +12:27:07.7  &     49$\pm$4 NE &    160$\pm$4 SW & - & 12.60(2,1) &   \\
SDSSJ1353+1138 & 13:53:06.34 +11:38:04.7  &     79$\pm$4 N &     32$\pm$4 S & - & 12.9(2,2) &   \\
ULASJ1405+0959 & 14:05:15.44 +09:59:29.8  &     37$\pm$4 E &     48$\pm$4 N &     23$\pm$4 S & - &   \\
SDSSJ1442+4055 & 14:42:54.70 +40:55:35.6  &     64$\pm$5 E &     38$\pm$5 W & - & - &   \\
SDSSJ1452+4224 & 14:52:11.50 +42:24:29.6 &      &      &     & - &   \\
SDSSJ1455+1447 & 14:55:01.88 +14:47:34.8  &     63$\pm$4  & -  & - & 12.6(3,2) &   \\
SDSSJ1458$-$0202 & 14:58:47.62 $-$02:02:05.3 &      &      &     & - &   \\ \hline
\end{tabular}
\end{table*}

\begin{table*}
\contcaption{}
\begin{tabular}{lcccccl}\hline
Source name & Coordinates & Flux density & Flux density &Flux density & $\log_{\rm 10}(L_{\rm FIR}/L_{\odot})$ & Comments \\
&(RA, Dec J2000)&(cpt 1,$\mu$Jy)&(cpt 2,$\mu$Jy)&(cpt 3,$\mu$Jy)&&\\ \hline
SDSSJ1515+1511 & 15:15:38.54 +15:11:35.1  &     70$\pm$12 E &     61$\pm$12 W & - & - &   \\
ULASJ1527+0141 & 15:27:20.22 +01:41:40.1 &      &      &     & - &   \\
ULASJ1529+1038 & 15:29:38.89 +10:38:04.3 &      &      &     & - &   \\
    J1616+1415 & 16:16:46.42 +14:15:43.9 &      &      &     & - &   \\
SDSSJ1620+1203 & 16:20:26.23 +12:03:40.7  &    327$\pm$6  & -  & - & $<$12.00 & LG,$\alpha=-0.1\pm0.1$ \\
    J1623+7533 & 16:23:16.92 +75:33:17.3 &      &      &     & - &   \\
  PSJ1640+1045 & 16:40:18.17 +10:45:05.4 &      &      &     & - &   \\
  PSJ1831+5447 & 18:31:27.12 +54:47:59.6  &    777$\pm$9 SE &    553$\pm$9 C &   2429$\pm$9 NW& - & 2im+LG,$\alpha_{\rm im}=-1.9\pm0.1$ \\
    J1949+7732 & 19:49:36.28 +77:32:39.0  &     39$\pm$5 E &     50$\pm$5 W & - & - &   \\
  PSJ2124+1632 & 21:24:16.85 +16:32:17.2  &     27$\pm$4 N &     22$\pm$4 S & - & - &   \\
 HE2149$-$2745 & 21:52:07.46 $-$27:31:49.4  &     45$\pm$5  & -  & - & 12.90(1,2) &   \\
   HS2209+1914 & 22:11:30.30 +19:29:12.8  &    413$\pm$8 S &    507$\pm$8 N & - & - & $\alpha=-1.1\pm0.1$ \\
  A2213$-$2652 & 22:13:38.38 $-$26:52:27.1  &     25$\pm$5  & -  & - & - & M \\
    J2250+2117 & 22:50:34.49 +21:17:24.0  &    193$\pm$19  & -  & - & - & LG;$\alpha=0.22\pm0.15$ \\
    J2257+2349 & 22:57:25.37 +23:49:30.4 &      &      &     & - &   \\
  PSJ2305+3714 & 23:05:55.78 +37:14:20.8  &     23$\pm$4 E &     31$\pm$4 W & - & - &   \\
  PSS2322+1944 & 23:22:07.16 +19:44:23.0  &     16$\pm$4 S &     38$\pm$4 N & - & 13.58(01,01) &   \\
WISE2329$-$1258 & 23:29:57.84 $-$12:58:59.0  &    287$\pm$18 NE  &    161$\pm$12 N  &    280$\pm$12 S & - &   \\
PSJ2332$-$1852 & 23:32:19.32 $-$18:52:06.6  &     59$\pm$11 E &    125$\pm$11 C &     59$\pm$11 W& - & 2im+LG,$\alpha=-0.8\pm1.0$ \\
ULASJ2343$-$0050 & 23:43:11.94 $-$00:50:34.3  &     36$\pm$4  & -  & - & $<$11.8 &   \\
    J2350+3654 & 23:50:07.54 +36:54:34.5 &      &      &     & - &   \\ \hline
\end{tabular}

\end{table*}

\begin{figure*}
\includegraphics[width=17.9cm]{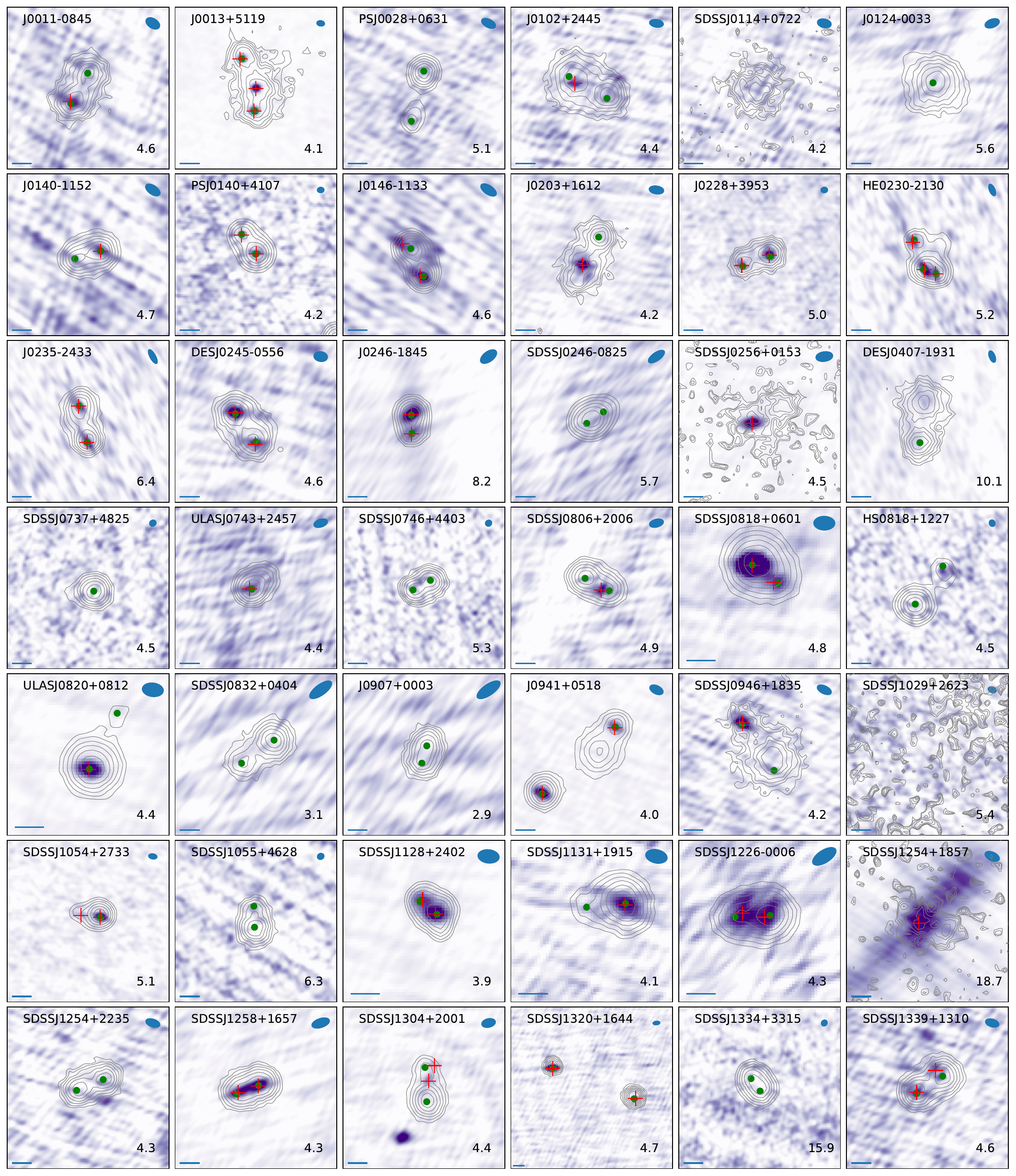}
\caption{VLA radio images (purple) and Pan-STARRS optical images (contours) of the sample. The r.m.s. noise, $\sigma$, in the radio images is indicated (in $\mu$Jy\,beam$^{-1}$) in the bottom right corner, and the flux scale runs from $-1\sigma$ to 5$\sigma$ or 70\% of the maximum, whichever is greater (30\% for SDSS\,J0818+0601, SDSS\,J1320+1644, SDSS\,J1349+1227 and PS\,J1831+5447). Contours in the optical images begin at 1/8 of the maximum brightness and increase in multiples of $\sqrt{2}$. The bar in the bottom left corner of each plot represents 1 arcsecond. The CLEAN beam is reproduced at the top right of each panel. Maps are centred at the coordinates given in Table~\ref{obstable}. Crosses indicate radio components which have been identified and fitted. {\it Gaia} point sources are identified by green blobs, and the Pan-STARRS maps have been re-centred by eye, typically by 100-200\,mas, to agree with the {\it Gaia} (and radio) astrometric frame.}
\label{images}
\end{figure*}

\setcounter{figure}{1}
\begin{figure*}
\includegraphics[width=17.9cm]{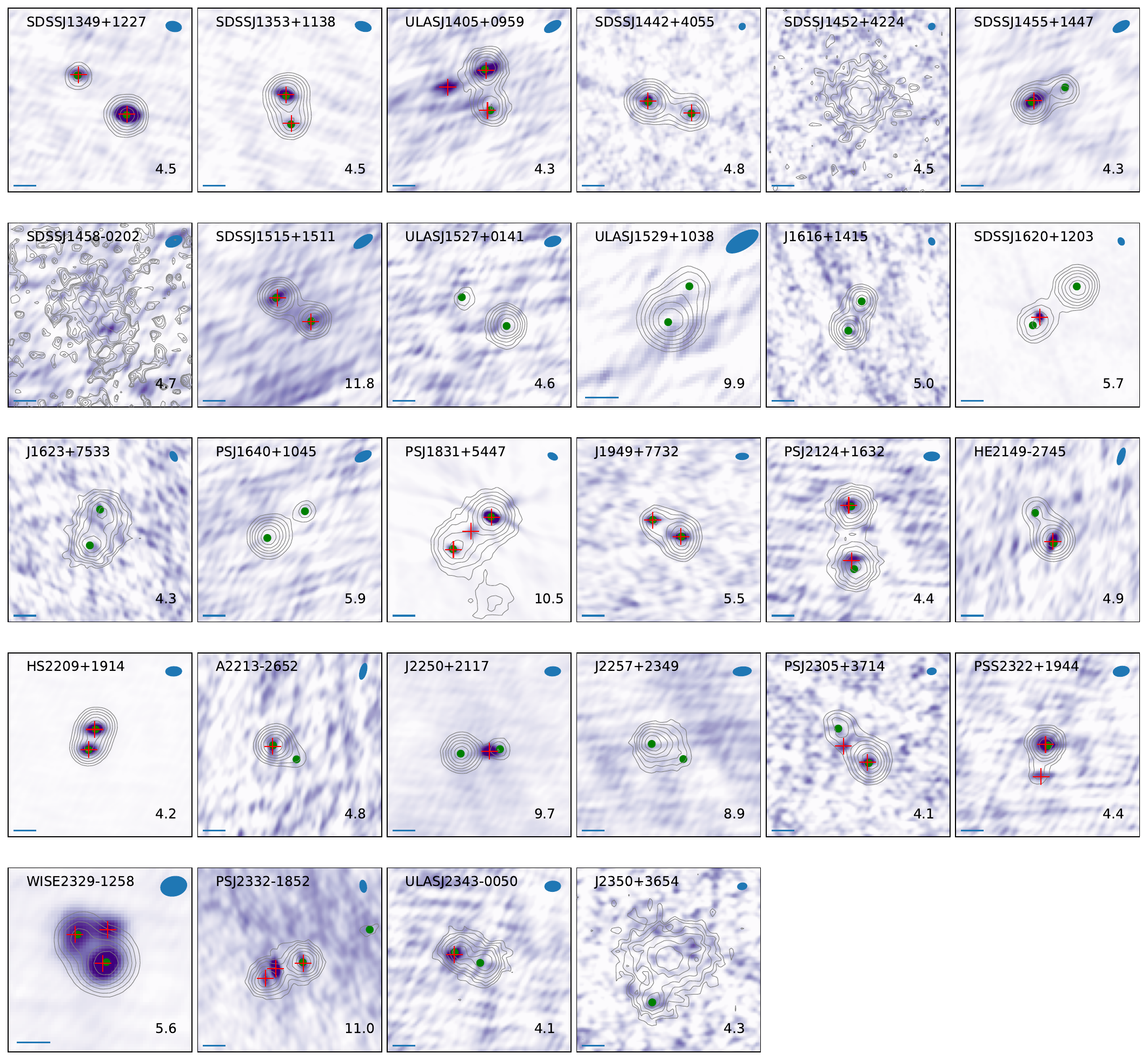}
\caption{continued}
\end{figure*}

\subsection{Notes on individual objects}

In the majority of cases, there is either no radio detection of any components of the lens system, or both the images of the lensed radio source are detected (Table~\ref{obstable}). We comment briefly on systems in which the identification of radio components are in doubt, or where their origin is not obvious from the imaging presented in Fig.~\ref{images}.

\subsubsection{Marginal detections}

 In seven cases (J0011$-$0845, J0102+2445, J0140$-$1152, PSJ\,0140+4107, ULAS\,J0743+2457, SDSS\,J1339+1310 and A2231$-$2652) we have detections of radio flux \change{density} at one or more places within the source, the brightest of which is within 1$\sigma$ of 20\,$\mu$Jy; such detections are therefore marginal or untrustworthy. 
 In the case of J0102+2445, the LOFAR DR2 survey image \citep{2022A&A...659A...1S} shows a very marginal ($\sim2.7\sigma$) possible detection at the 400\,$\mu$Jy level. In PSJ\,0140+4107 the appearance of two separate components close to the expected positions, together with relatively low noise in the map, implies that the components are real.

\subsubsection{J0013+5119}

\cite{2019MNRAS.483.4242L} detect two lensed images and a lensing galaxy between them. In the VLA observations, we detect radio emission from both lensed images as well as the lensing galaxy.

\subsubsection{J0146$-$1133}

\cite{2018MNRAS.479.5060L} detected two lensed images in the optical, approximately equal in brightness, together with a faint lensing galaxy very close to the northern component. These lensed images have a separation of 1\farcs69 \citep{2018MNRAS.479.5060L}, with an error probably of order 0\farcs01. Our fitting gives a separation of 2\farcs01$\pm$0\farcs09, a significant difference from the optical separation, but the two radio components also appear oriented at a different angle to the line joining the optical components. There is no obvious explanation for this difference, other than one of the radio components being spurious (which is possible due to its relative faintness).

\subsubsection{J0203+1612}

This system is listed as a ``probable lens'' by \cite{2019MNRAS.483.4242L} as it does not have final spectroscopic confirmation. We are also unable to confirm it as we do not detect radio emission from either of the potentially lensed components. Instead, we detect a 46$\pm$4\,$\mu$Jy source, which appears slightly extended, at the position of the proposed lensing galaxy.

\subsubsection{HE\,0230$-$2130}

\cite{1999A&A...348L..41W} discovered this lens system, which is one of the few objects in the sample that is not a double-image lens system. It has two lensing galaxies, and the resulting complex Fermat surface would be expected to result in five images; one is not observed, possibly due to a dark-matter subhalo \citep{2023arXiv230805181E}. We detect emission from three images, including the two bright merging images A and B. These are fitted separately, together with a third component to represent image C. 

\subsubsection{DES\,J0245$-$0556}

There are two detections of radio components, which are roughly coincident with lensed images seen in the Pan-STARRS survey. The radio and optical separations are consistent; the apparently slightly greater radio separation is due to the lensing galaxy, which appears blended with one of the optical components \citep{2018MNRAS.479.4345A,2021MNRAS.503.1557S}.

\subsubsection{SDSS\,J0246$-$0825}

We detect no radio emission from this object. A high-resolution Keck optical image is available \citep{2021MNRAS.503.1557S} which shows the two quasar images at the {\it Gaia} positions, together with the lensing galaxy slightly north of the line between them.

\subsubsection{SDSS\,J0256+0153}

Optical imaging \citep{2016MNRAS.456.1595M} shows two lensed components together with a more diffuse lensing galaxy. We clearly detect one radio component at 114$\pm$5\,$\mu$Jy, close to the brighter A component. However, astrometry conducted by \cite{2016MNRAS.456.1595M} places the lensing galaxy only 0\farcs5 from A, with the B component being only about 0.3-0.4 magnitudes fainter than A. Since we detect no other radio component at the ratio of at least 5:1, and given the astrometric errors, it is likely that the radio detection is in fact emission from the lensing galaxy.

\subsubsection{DES\,J0407$-$1931}

The noise level in this radio map is approximately a factor of 2 higher than most of the other maps \change{(about 10\,$\mu$Jy compared to the typical 5\,$\mu$Jy)}, due to the difficulty in subtracting a nearby bright source; there is a slight increase in the background flux close to one of the optical images, but this is unlikely to be a genuine detection.

\subsubsection{ULAS\,J0743+2457}

This object is detected only marginally above the 20-$\mu$\change{Jy beam$^{-1}$} threshold, in an image with slightly raised noise levels due to residuals associated with nearby sources. Its position is coincident with the brighter component of an optical double lens system. Adaptive optics imaging \citep{2016MNRAS.458....2R} shows the two separate resolved components, with the much weaker one very close to the quasar.

\subsubsection{SDSS\,J0806+2006}

VLT and Keck adaptive optics imaging in the near-infrared \citep{2008A&A...492L..39S,2021MNRAS.503.1557S} shows two images of the lensed quasar, with the brighter one to the northeast, approximately 0.8 magnitudes brighter at 1.6\,$\mu$m than the fainter one. The lensing galaxy lies close to the fainter image. Since our radio detection is also very close to the fainter infrared image, it is likely to be a detection of radio emission from the lensing galaxy.

\subsubsection{SDSS\,J0818+0601}

\cite{2016MNRAS.456.1595M} refer to this object as a possible quasar pair, as the lens galaxy was not detected, but later spectroscopy \citep{2020A&A...633A.101H} confirmed its status as a double-image lens system, as well as detecting the presence of microlensing. The optical flux ratio is approximately 6:1, slightly greater than the fitted ratio of the two radio components detected.

\subsubsection{ULAS\,J0820+0812}

This lens system has a 2\farcs3 separation between the lensed images \citep{2008MNRAS.387..741J} and a high (6:1) optical flux ratio, with the fainter object just visible to the northwest in Fig.~\ref{images}. Higher-resolution imaging by \cite{2013ApJ...765..139R} shows the lensing galaxy closer to the faint component. The coincidence of the radio detection with the brighter component on the Pan-STARRS image strongly implies that the radio emission comes from the lensed object, with the other lensed image not detected (its expected flux density would be about 15\,$\mu$Jy). Fitting an extended Gaussian to the radio detection, instead of a point source model, gives an upper limit of 0\farcs3 (2.5\,kpc) on the size of the source.

\subsubsection{J0941+0518}

Optical imaging reveals a wide-separation (5\farcs4) lens system with a relatively bright lensing galaxy. In the VLA image, we clearly detect both lensed images and do not detect the lensing galaxy. The in-band spectral index, derived from splitting the band in two, is relatively flat (Section~\ref{subsection:spectral}).

\subsubsection{SDSS\,J0946+1835}

Optical/NIR imaging of this system \citep{2016MNRAS.458....2R} shows two condensations, the northern one containing image A of the lensed quasar, and the southern one containing image B and the lens galaxy, with image B being 0.47 magnitudes fainter in $K^{\prime}$ (2.1\,$\mu$m). We detect image A at 33$\pm$4\,$\mu$Jy, but even a detection limit of 20\,$\mu$Jy in the VLA images implies a larger A/B flux ratio in the radio than the near-infrared. Possible explanations include microlensing at the shorter wavelengths, or variability coupled with time delay effects.

\subsubsection{SDSS\,J1054+2733}

Two radio components are detected here, which are consistent with being coincident with the two lensed images \citep{2010AJ....139.1614K}.

\subsubsection{SDSS\,J1254+1857}

Mapping this object is relatively difficult because of the presence of a brighter, 75-mJy source about 2 arcmin away. Although at least one radio component is detected, the noise is high and it is difficult to evaluate whether a second is present. The detected component may be the lensing galaxy since the quasar images differ by only 0.24 magnitude in flux \citep{2016MNRAS.456.1595M}.

\subsubsection{SDSS\,J1304+2001}

There are two optical condensations detected in optical imaging associated with the discovery paper for this lens system \citep{2010AJ....139.1614K}, conducted with the University of Hawaii 2.2-m and Subaru telescopes. The southern condensation consists of a nearby galaxy G2, which is detected as a radio source here. The northern condensation consists of two quasar images and a galaxy G1. G1 lies between the lensed images and is about 3\farcs5 from G2.  We detect the presence of a faint radio component between the two quasar images, which is likely to be the galaxy G1. There is a further possible radio component to the north of this, and slightly to the west of the fainter quasar image, whose presence we are unable to explain \change{via lensing}.

\subsubsection{SDSS\,J1320+1644}

This object is a large-separation (8\farcs6) double-image lens system \citep{2013ApJ...765..139R}. There are two primary lensing galaxies, situated either side of the line between the lensed components; neither are detected in the radio. The noteworthy feature of this system is the extreme discrepancy in the flux ratio between the two lensed images; the western image is denoted as A by \cite{2013ApJ...765..139R} as it is generally brighter in the optical, typically by 0.3--0.4 magnitudes. However, the eastern image is a factor of 5.6 times brighter in the radio, with the western image barely detectable. Such a discrepancy requires either extreme variability or, more likely, a significant change in the optical fluxes by the presence of microlensing or extinction. This is discussed further in Section~\ref{section:discussion}.

\subsubsection{ULAS\,J1405+0959}

We detect two radio components in this system, which are almost certainly lensed images of the background quasar. We also detect an extra radio component, to the east of the main north-south axis of the lensed images. This coincides \citep{2012MNRAS.419.2014J} with a very red object seen in the data from the UKIRT Infrared Deep Sky Survey (UKIDSS, \citealt{2007MNRAS.379.1599L}), which is probably a nearby galaxy, the lensing galaxy being close to one of the lensed images \citep{2014AJ....147..153I}. 

\subsubsection{SDSS\,J1458$-$0202}

This lens system was identified by \cite{2016MNRAS.456.1595M} as a doubly-imaged quasar with separation 2\farcs1 along a NE-SW axis. The lensing galaxy is in between the lensed images, and is diffuse and dominates the Pan-STARRS map. We do not detect any radio flux from this object.

\subsubsection{SDSS\,J1620+1203} 

\cite{2010AJ....139.1614K} discovered this lens system and detected the lensing galaxy close to the fainter (southeastern) optical image. The single radio component detected in these observations is also very close to the fainter optical image and is, therefore, likely to originate in the lensing galaxy, leaving the lensed quasar images undetected. The alternative explanation, that there is an extreme difference in flux ratio between the optical and radio, appears less likely.

\subsubsection{J2250+2117}

The radio flux detected in this system is almost certainly from the lensing galaxy, which is close to the weaker western lensed image \citep{2019MNRAS.483.4242L}. There is radio flux at about the 6$\sigma$ level ($\sim 600\,\mu$Jy\,beam$^{-1}$) in LOFAR-DR2 \citep{2022A&A...659A...1S} at 150\,MHz.

\subsubsection{PS\,J2305+3714}

We detect two radio condensations in this system, one of which is coincident with the brighter optical lensed image. Given the $\sim$1\,mag difference in the optical fluxes of the images \citep{2018MNRAS.479.5060L}, we would not expect to detect the fainter optical image. We also marginally detect a second radio component, which may be associated with the lensing galaxy given the coincidence with the position from \cite{2021MNRAS.503.1557S}, or may be spurious since it is very close to our detection threshold of 20\,$\mu$Jy\,beam$^{-1}$.

\subsubsection{WISE\,2329$-$1258}

WISE\,2329$-$1258 is clearly detected in these observations, and has a previous detection in the NRAO-VLA Sky Survey with a flux density of $2.4\pm0.5$\,mJy at 1.4\,GHz \citep{1998AJ....115.1693C}. The radio and optical image overlay in Fig.~\ref{images}  is based on the {\it Gaia} astrometry, from which it appears that the brightest radio and optical components are coincident. Therefore, in principle, this object is a double-image lens system in which the lensed quasar images are visible both in the optical and radio. However, there is a clear detection of a third radio component, about 1\arcsec\ north, and slightly west, of the brightest component. 

\begin{figure*}
\begin{tabular}{cc}
\includegraphics[width=7cm]{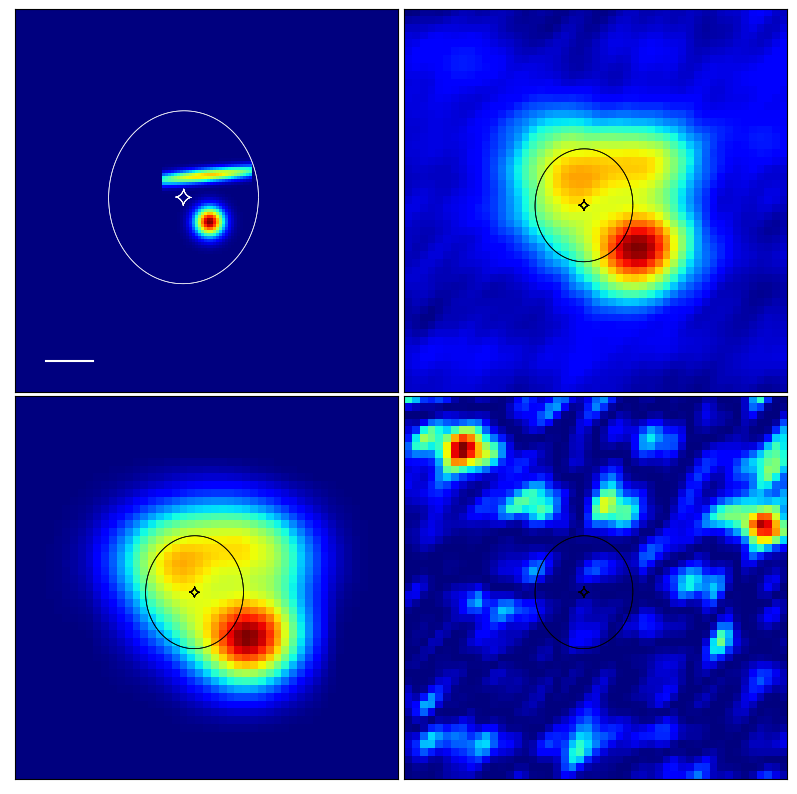}&
\includegraphics[width=10cm]{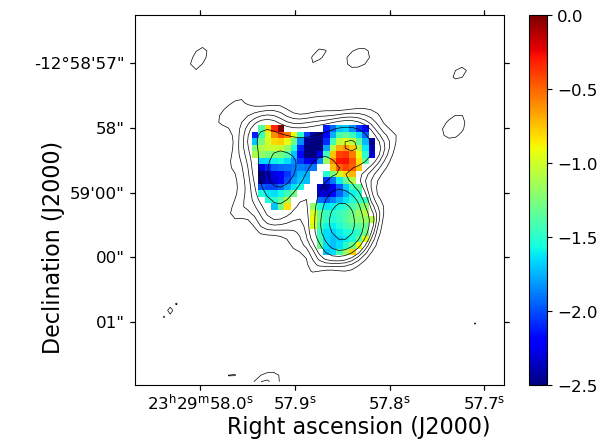}\\
\end{tabular}
    \caption{Left, top row: Reconstructed source plane, using a singular isothermal sphere lens model, and observed map of WISE\,2329$-$1258. Bottom row: predicted image plane using the singular isothermal sphere model, and chi-squared residual. Right: In-band spectral index image of WISE\,2329$-$1258 constructed from images at 4-6\,GHz and 6-8\,GHz, with the higher-frequency image tapered at 400\,k$\lambda$ to approximately match the resolution of the lower-frequency image and restored with the same 0\farcs6 resolution {\sc clean} beam. 
    }
    \label{wise2329}
\end{figure*}

A Keck adaptive-optics image of this source \citep{2021MNRAS.503.1557S} shows the two arcsecond-scale optical images of the background quasar, together with the lensing galaxy close to the fainter, north-eastern component. We have used the separation between the bright image and the lensing galaxy from the Keck observation to set the position of the lensing galaxy in the radio image, assuming that the brightest points of the radio and optical emission are coincident. This has been adjusted by hand using a singular isothermal sphere model for the lensing galaxy until the brightest radio image is satisfactorily reproduced. An extra radio component is then added to the source plane, represented by a Gaussian in the source plane, and its properties, together with all parameters of the lensing galaxy except the position, are then allowed to vary. The best-fitting source model is shown in Fig.~\ref{wise2329}. The second fitted component is implied to be extremely elliptical, in a direction nearly perpendicular to the vector to the first component, and the isothermal galaxy model becomes mildly elliptical. However, the available data do not constrain the parameters of the extra source component or the lensing galaxy very well, apart from the position angle and ellipticity of the extra radio source component. Further higher-resolution radio data is needed to elucidate the structure of the source further.

\subsubsection{ULAS\,J2343$-$0050}

This double-image lens system \citep{2008MNRAS.387..741J} has an image separation of 1\farcs4, with the brighter optical/NIR image (by 0.2 magnitudes in $R$ and 0.6 in $g$) at the western end. The lens galaxy is closer to the brighter, western image. Here we have a single radio detection, which is likely to correspond to the eastern component, with possibly a small hint of some radio flux further west. The radio flux ratio, with the detected radio component a factor of $\simeq2$ above the detection threshold, is therefore significantly different from the optical flux ratio.

\section{Discussion}
\label{section:discussion}

\subsection{Radio in-band spectral indices}
\label{subsection:spectral}

In a minority of cases, where the radio source is relatively strong, we can split the wide radio band into two parts (4-6\,GHz and 6-8\,GHz) and attempt to calculate an in-band radio spectral index. In this process, the higher-frequency dataset is imaged using a Gaussian taper of width 350\,k$\lambda$ to weight down the high $u,v$ end of the dataset and restored with the same restoring beam as the lower-frequency dataset. Spectral indices, where available, are included in Table~\ref{obstable}. This is interesting because it provides an additional diagnostic of the radio emission physics. Synchrotron emission from either an extended jet component of an AGN or from synchrotron electrons associated with star-forming regions should have a relatively steep spectrum; emission from an optically thick AGN core or small corona \citep{Laor2008} should have a flatter spectrum.

Despite high noise due to the limited spectral range, the majority of lensed images \change{(6/7 cases)} appear to have relatively steep spectra (generally $-1.2<\alpha<-0.7$), suggesting the presence of lensed synchrotron emission from the background radio source associated with the quasar. Some spectral indices appear extremely steep, despite the efforts to image with the same $uv$ plane weighting at both frequencies. The single exception is J0941+0518, which is consistent with a flat radio spectrum, within the errors. On the other hand, the three lensing galaxies, which are bright enough to attempt to derive spectral indices, show relatively flat spectra consistent with measurements of 150-1400\,MHz spectral indices in more nearby early-type galaxies that are compact on LOFAR scales of 6$^{\prime\prime}$ \citep[][and references therein]{2022A&A...660A..93C, 2023A&ARv..31....3B}.



\subsection{Optical and radio flux ratios}

\begin{figure}
\includegraphics[width=8cm]{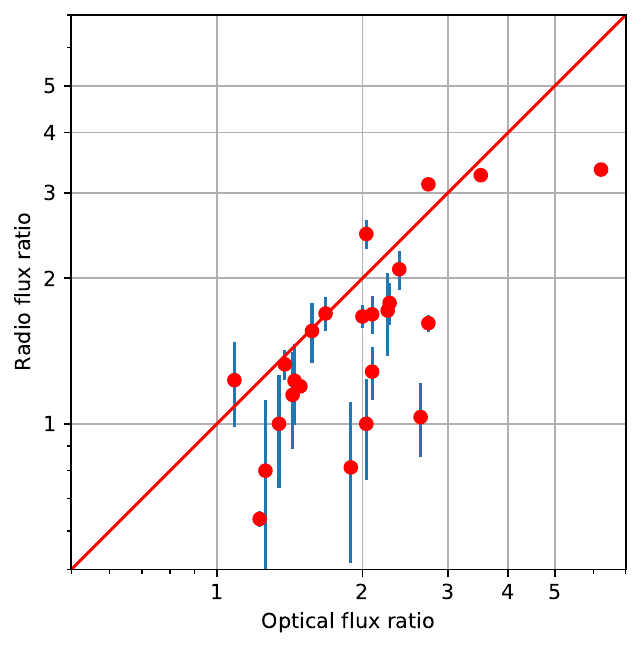}
\caption{Optical and radio flux ratios for the lensed quasars in the sample, for the objects in which two radio components are detected, and excluding those for which the radio components are likely to originate in the lensing galaxy. The line represents equality between the fluxes at the two wavelengths. Optical flux ratios are taken from Gaia data releases \protect\citep{2018A&A...616A...1G} and ratios are defined such that the optical flux ratio is $>$1. SDSS\,J1320+1644 (2.11, 0.18) is not shown.  }
\label{ratio}
\end{figure}

In Fig.~\ref{ratio} we plot the radio and optical flux ratios between the two lensed images of objects in the sample. The optical flux ratios are taken from the measurements of the compact sources in GAIA \citep{2018A&A...616A...1G}, and the 6-GHz radio fluxes from this work; objects with significant radio emission from the lensing galaxy, and WISE\,2329$-$1258, have been excluded. Most of the ratios are within 3 standard deviations of the equality line, although the errors are frequently large for the fainter radio sources. There is a tendency for the radio flux ratios to be slightly lower than the optical ones, the most notable case being the most asymmetric double lens in the sample, SDSS\,J1320+1644, which is one of four objects in which the brighter optical image corresponds to the fainter radio one.

The primary influence on the flux densities of images in gravitational lens systems is the source position and structure, combined with the lens mass macromodel and any structure within it. Flux densities in lensed images may vary with wavelength if the source has different structures at different wavelengths, with structure close to caustics being more highly magnified \citep[e.g.][]{2000ApJ...535..692K,2009MNRAS.394..174M,2012MNRAS.424.2429S}. Even relatively simple sources can suffer significant differential magnification if the lens mass distribution has structure on smaller scales. Lens models which include  $10^6-10^9M_{\odot}$ substructures on top of the macromodel produce flux anomalies; that is, image flux ratios which differ from that predicted by the macro model \citep{2002ApJ...580..685S,2021MNRAS.503.1557S}. The superposition of finite-size sources on magnification patterns produced by the substructures can result in image flux densities which are sensitive to the size of the source \citep{2006MNRAS.365.1243D} and affect both radio and optical lensed images, though not equally due to the larger size of the radio source. Microlensing by stars in the lensing galaxy produces large independent changes in the brightness of different lensed images, with optical fluxes being exclusively affected, because the small size of the optical source is matched to the characteristic scale of the caustic structures corresponding to a set of  $\sim 1M_{\odot}$ \change{objects \citep{1979Natur.282..561C,1989AJ.....98.1989I}; see \cite{2024SSRv..220...14V} for a recent review}.

Two other effects can also contribute to different fluxes in lensed images. First, source variability combined with the differential time delay between images can produce different flux ratios, with likely higher amplitude of variation in the optical where physically smaller parts of the background quasar source dominate. Second, extinction in the lens galaxy can produce chromatic effects \citep[e.g.][]{1997A&A...317L..39J,2000MNRAS.311..389J}, with short-wavelength fluxes and fluxes of the fainter image in double systems, which are closer to the line of sight to the lensing galaxy, being preferentially suppressed. For example, \cite{2006ApJS..166..443E} found evidence for differential extinction of lensed images in the lensing galaxy, with typical $A_V\sim0.56$, in the majority of a sample of 10 lensing galaxies.

A number of the systems in this work have previously been studied at multiple wavelengths in the optical/IR to understand the image flux ratios. In SDSS\,J1515+1511, for example, there is likely to be a complex interplay of microlensing and extinction effects taking place \citep{2017ApJ...836...14S}. Investigation of the cores of emission lines, corresponding to emission from slowly moving gas occupying more extended regions which are relatively immune to microlensing, suggests a modest visual extinction ratio between A and B of 0.13\,magnitudes. In SDSS\,J1339+1310 \citep{2014A&A...568A.116S,2016A&A...596A..77G} the microlensing/extinction separation is complicated by the line emission being possibly affected by microlensing, but modest inter-component extinction ratios are again implied by modelling.

In this sample, we see a mixture of large discrepancies between optical and radio flux ratios of the lensed images and a general trend for slightly larger optical ratios, typically of the order of 10 percent. Previous detailed attempts to disentangle effects on flux ratios (e.g. \citealt{2012MNRAS.419..936F,2015MNRAS.454..287J}) have used some combinations of substructures, finite sources, microlensing and extinction. Here, it is likely that microlensing is responsible for some of the extreme cases of different optical flux ratios, with substructure/finite source effects broadening the distribution and an overall slight, but systematic, raising of the optical flux ratio due to differential extinction in the optical which affects the flux ratios of the lensed images such that the fainter image, being closer to the line of sight to the lensing galaxy, is slightly demagnified.


\subsection{Far-infrared/Radio Correlation}
The far-infrared--radio correlation (FIRC) is a correlation between radio and far-infrared luminosity which is observed for galaxies over a range of orders of magnitude in luminosity and star-formation rate \citep{1991MNRAS.251P..14S}. It arises physically because star-forming processes give rise both to FIR emission from heated dust, and to radio emission via acceleration of electrons in supernova remnants. We can parameterise the relation of IR and radio luminosity using the $q_{\rm IR}$ parameter, defined as

\begin{center}
   \begin{equation}
    q_{\rm IR} = \rm{log_{10}}\left (  \frac{\textit{L}_{IR}}{3.75\times10^{12}\textit{L}_{1.4\rm GHz}}\right ),
    \label{QIR}
\end{equation} 
\end{center}

\noindent where $L_{\rm IR}$ is the total integrated infrared luminosity from 8-1000\,$\mu$m in the emitted frame, and $L_{1.4\rm GHz}$ is the luminosity at 1.4\,GHz, in W\,Hz$^{-1}$.  \change{$L_{1.4\rm GHz}$ is calculated from the flux density $f_6$ in W\,m$^{-2}$\,Hz$^{-1}$, measured at 6\,GHz, assuming a spectral index of $\alpha=-0.7$ for the K-correction):}

\change{
\begin{equation}
    L_{1.4\rm GHz} = \left(\frac{6}{1.4}\right)^{-\alpha}\frac{4\pi D_L^2}{(1+z)^{1+\alpha}}\ f_6,
\end{equation}
}

\noindent \change{where $D_L$ is the luminosity distance, and the factor on the left corrects from emitted luminosity at 6\,GHz to that at 1.4\,GHz.} Galaxies lie on the main correlation line corresponding to star-forming galaxies if their $q_{\rm IR}=2.40\pm 0.24$ \citep{IvisonFIRC}. Galaxies with $q_{\rm IR}<2.40$\ have excess radio emission which is likely to originate in an AGN \citep{Peterson1997,Heckman2014}.

\cite{2018MNRAS.476.5075S} previously studied this correlation by observing a large sample of known lensed quasar systems with {\it Herschel} to derive FIR luminosities, $L_{\rm FIR}$, obtained by integrating the implied rest-frame flux density from $40\,\mu$m to $120\,\mu$m \citep{Helou1988}. Upper limits (in all cases) are also available from the AKARI far-infrared all-sky survey \citep{2015PASJ...67...50D} and the IRAS Point Source Catalogue  \citep{1994yCat.2125....0J}, but the 5$\sigma$ detection level of these surveys typically correspond to $L_{\rm FIR}\sim 10^{14}L_{\odot}$, which does not constrain the position of the radio source on the radio - far-infrared correlation for all but the strongest radio sources. We have therefore used flux densities \change{and $L_{\rm FIR}$ values} from \cite{2018MNRAS.476.5075S} only. FIR luminosities may then be converted to $L_{\rm IR}$ by multiplying by 1.91 \citep{2001ApJ...549..215D,
2018MNRAS.476.5075S}.





\begin{figure*}
\includegraphics[width=\textwidth]{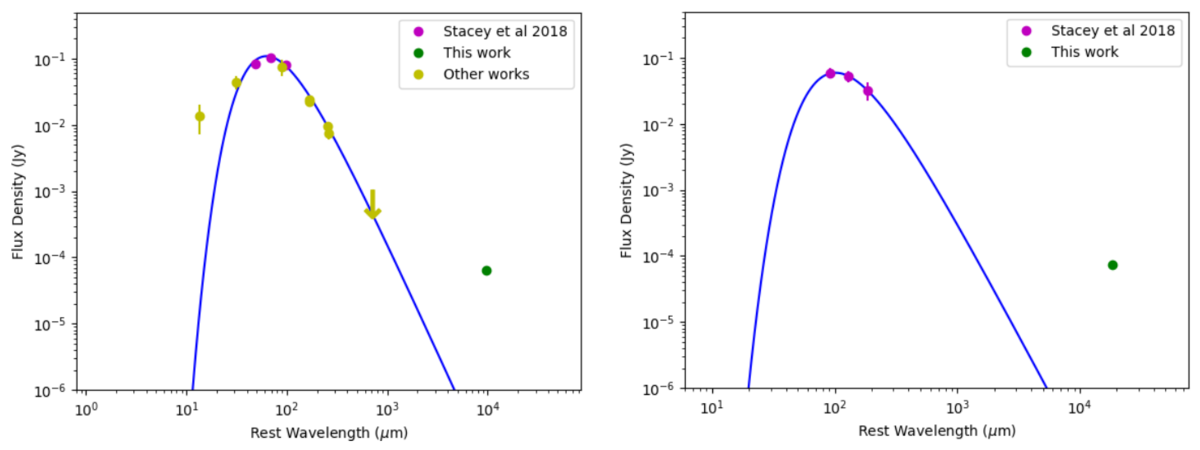}\vspace{-10pt}
\caption{Spectral energy distributions of two of the sample, with modified blackbody fits. Left: PSS\,J2322+1944, where a wide range of literature data is available \citep{2001A&A...374..371O,2001ApJ...555..625C,2002MNRAS.329..149I,2002A&A...387..406C,2017arXiv170505693M,2018MNRAS.476.5075S}; right: SDSS\,J1349+1227 \citep{2018MNRAS.476.5075S}.} 
\label{irsed}
\end{figure*}

\begin{figure}
    \centering
    \includegraphics[width=95mm,scale=0.5]{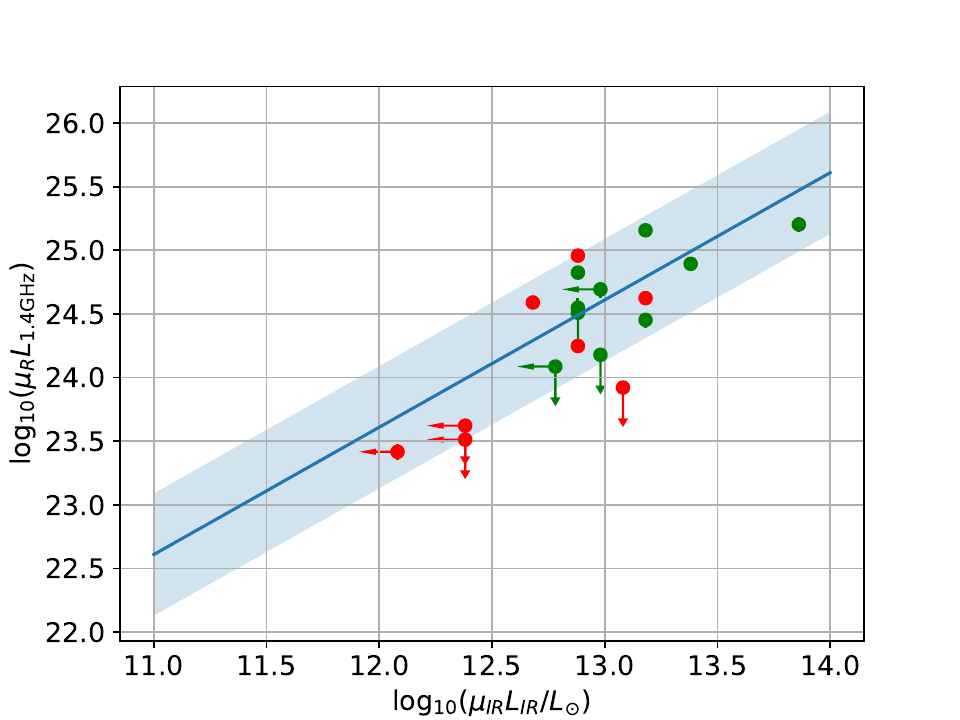}
\caption{FIR-radio correlation for the securely detected radio sources. The blue stripe and line represent the FIR-radio correlation together with the $\pm 2\sigma$ scatter which are taken from Ivison et al. (2010). Systems with sources of redshift below and above 1.8 are plotted with red and green symbols, respectively. }
    \label{FIRC}                               
\end{figure}

Fig.~\ref{irsed} shows a sample of spectral energy distributions derived from the radio and available infrared fluxes in some of the objects in this sample, and Fig.~\ref{FIRC} shows the radio-infrared correlation derived from them, together with the correlation derived from star-forming galaxies \citep{IvisonFIRC}. All of the objects for which FIR fluxes exist are on the FIRC to within 2$\sigma$, except for SDSS\,J0246$-$0825 which is slightly below. There is therefore no evidence for radio excess in any of the objects in the subsample with FIR information. 

Both of the axes of Fig.~\ref{FIRC} show luminosities which have been boosted by a magnification factor due to the lensing; we have assumed in the analysis that the boosting factor $\mu_{\rm IR}$ in the far infra-red is equal to the factor $\mu_{\rm R}$ in the radio. This is likely to be true for star-forming objects, where the radio emission and heated dust originate in the same region. \change{In the case of AGN-related radio emission, however,  any spatial offset between AGN and star-forming regions could result in unequal magnification factors.  Observations of the lensed radio-loud AGN JVAS B1938+666, for example, find a radio-to-infrared magnification factor ratio of about 10 \citep{10.1046/j.1365-8711.1998.295241.x,Barvainis_2002,Lagattuta_2012}. Further, if positioned near a lensing caustic curve, a more compact source will undergo a greater magnification than a coincident extended source \citep{2012MNRAS.424.2429S,10.1093/mnras/sty513}. Given that all but one of our radio sample has double images, which are formed outside the very high magnification region near the tangential caustic, the radio emission is unlikely to be significantly boosted with respect to the infrared \citep{Hezaveh_2012}. On the other hand, the infrared emission could be moderately boosted with respect to the radio should an offset infrared source lie close to the tangential caustic, particularly in the position inside the caustic cusp. There is a small possibility, therefore, that the FIR-radio ratio of some of our objects is slightly overestimated.    }

\change{Assuming a typical lensing magnification factor of around $\mu \sim$ 5-10 for this sample, our $q_{\rm IR}$ results probe the quasar radio luminosity function between around $L_{\rm 1.4 GHz}\sim10^{23}$--$10^{24.5}$\,W\,Hz$^{-1}$ \citep{2021ApJS..255...30G} across a redshift range 0.79--4.82 with median 1.8.   Our findings differ from those of \cite{White_2017} who, within a narrow redshift window of  $0.9\lt z \lt 1.1$, find a radio excess from the FIRC parameter $q_{125}$ as calculated using the monochromatic rest-frame 125 \textmu m luminosity. \cite{2023arXiv231210177C} also find a radio excess from the FIRC, determined using the total infrared luminosity, in a large sample of AGN from the LoTSS Deep Field survey at 150 MHz. We caution, however, that known radio-loud objects were specifically excluded here from the input observing catalogue; in particular, 28 sources originating in radio-selected lens detection programmes that are mostly CLASS sources \citep{2003MNRAS.341....1M}. Therefore, we are exploring the lower envelope of the distribution in Fig.~\ref{FIRC}.}

\change{The radio-detected sample used by \cite{White_2017} covers a slightly higher luminosity range than our sample, at $L_{\rm 1.5 GHz}\sim10^{23.5}$--$10^{24.75}$\,W\,Hz$^{-1}$. The sample of \cite{2023arXiv231210177C} again covers a higher luminosity range, with most objects in the redshift range covered by our sample above  $L_{\rm 1.4 GHz}\sim10^{24}$\,W\,Hz$^{-1}$, assuming a typical radio spectral index of $\alpha=-0.7$.  It is possible, therefore, that the results may reflect 
the extension by our study into a distinct, fainter, source population.
Indeed, \cite{2022MNRAS.515.5758M} use brightness temperature measurements made using the International LOFAR Telescope to suggest that two overlapping populations contribute to the observed radio luminosity distribution for radio-quiet AGN. On the other hand, evidence of jet activity in a $L_{\rm 1.6 GHz}\sim10^{22}$\,W\,Hz$^{-1}$ quasar at $z = 1.51$ \citep{2019MNRAS.485.3009H} -- which was found to lie within the scatter of the FIRC \citep{2018MNRAS.476.5075S} -- demonstrates that even relatively low-power radio emission can result from AGN activity and that the FIRC cannot always be used to rule out AGN activity.  Future deep surveys made by the Square Kilometre Array (SKA)-Mid telescope will extend the radio study of unlensed AGN down to the sub-\textmu Jy regime  \citep{braun2019anticipated}, allowing routine access to the $L_{\rm 1.4 GHz}\sim10^{22}$\,W\,Hz$^{-1}$ population at redshifts $z\sim1.5$ and the $L_{\rm 1.4 GHz}\sim10^{21.5}$\,W\,Hz$^{-1}$ population at redshifts $z\sim1$.}


\change{In principle, we can explore the evolution of the FIRC with redshift and stellar mass. However, there is no noticeable separation in $q_{\rm IR}$ between higher ($z>1.8$) and lower ($z<1.8$) redshift objects in our sample (plotted as different colours in Fig.~\ref{FIRC}). This is not very surprising; if the evolution of $q_{\rm IR}(z)$ is parametrized as $q_{{\rm IR},z=0}+\beta\log_{10}(1+z)$, there appears to be a mild decrease in $q_{\rm IR}$ with redshift, values of $\beta$=$-$0.04, $-$0.22 and $-$0.14 being found by \cite{2017MNRAS.469.3468C}, \cite{2022MNRAS.515.5758M} and \cite{2021A&A...647A.123D}, respectively. This is well within the scatter in Fig,~\ref{FIRC}; the same is true for stellar mass dependence \citep{2022A&A...668A..81D} in which each factor of 10 difference in stellar mass corresponds to a change of 0.148 in $q_{\rm IR}$. Modelling of the SED of the galaxy is challenging to this level of accuracy, given the proximity to the bright quasar component, with additional assumptions required in the case of lenses about the differential magnification of source and host.}

\section{Conclusions}

We present radio observations of 70 double-image gravitationally lensed systems, selected by optical flux, the vast majority of which do not have previous radio detections. We detect 40 of them down to a 4$\sigma$ limit of about 20\,$\mu$Jy, although 7 of these detections are marginal. Nevertheless, the median lensed radio flux density of this optically selected group of lensed radio-quiet quasars, at about $20\,\mu$Jy, corresponds to an intrinsic source flux density of a few $\mu$Jy.
The properties of these radio sources are, in most cases, consistent with their placement on the radio-FIR correlation and therefore consistent with the primary radio emission mechanism being star formation. However, the FIR data are incomplete for this sample, unlike previous investigations with Herschel data \citep{2018MNRAS.476.5075S}; caution is required both because of this and because weak radio sources may still harbour AGN emission \citep{2019MNRAS.485.3009H}. Most of the radio source spectral indices appear to be moderately steep, consistent with synchrotron emission associated either with AGN or with an origin in supernovae within star-forming regions; one lensed quasar (J0941+0518) has a flat spectrum consistent with a lensed, self-absorbed radio core.  

\change{We find that the} flux ratios of the lensed images in the radio correlate well with the optical flux ratios, with a hint of differences likely associated with mild extinction of the fainter components in the optical, together with larger discrepancies likely associated with optical microlensing.
Detection of radio \change{emission} from this sample is the first step in understanding the nature of radio emission in these radio-quiet quasars. Distinction between competing models of the radio emission requires investigation at higher radio resolution, to search for (or rule out) high brightness temperature emission from AGN. \change{The forthcoming SKA \citep{braun2019anticipated} will allow a much more detailed investigation of the faint end of the quasar luminosity distribution; given the detection rate in our VLA observations, we expect that the whole population of radio-quiet quasars should be detected with the SKA in the future.}

\section*{Acknowledgements}
We thank the referee for useful comments on the paper. JPM acknowledges support from the Netherlands Organization for Scientific Research (NWO) (Project No. 629.001.023) and the Chinese Academy of Sciences (CAS) (Project No. 114A11KYSB20170054). This work is based on the research supported in part by the National Research Foundation of South Africa (Grant Numbers: 128943)

The VLA is operated by the US National Radio Astronomy Observatory (NRAO), which is a facility of the National Science Foundation operated under cooperative agreement by Associated Universities, Inc. The Pan-STARRS1 Surveys (PS1) and the PS1 public science archive have been made possible through contributions by the Institute for Astronomy, the University of Hawaii, the Pan-STARRS Project Office, the Max-Planck Society and its participating institutes, the Max Planck Institute for Astronomy, Heidelberg and the Max Planck Institute for Extraterrestrial Physics, Garching, The Johns Hopkins University, Durham University, the University of Edinburgh, the Queen's University Belfast, the Harvard-Smithsonian Center for Astrophysics, the Las Cumbres Observatory Global Telescope Network Incorporated, the National Central University of Taiwan, the Space Telescope Science Institute, the National Aeronautics and Space Administration under Grant No. NNX08AR22G issued through the Planetary Science Division of the NASA Science Mission Directorate, the National Science Foundation Grant No. AST-1238877, the University of Maryland, Eotvos Lorand University (ELTE), the Los Alamos National Laboratory, and the Gordon and Betty Moore Foundation.

\section*{Data Availability}

VLA data gathered for this project are publicly available on the VLA archive, which may be accessed at {\tt https://data.nrao.edu/portal} under project numbers 20B-309 (PI Jackson) and 23A-278 (PI Jackson).



\bibliographystyle{mnras}
\bibliography{rqq} 
\bsp	
\label{lastpage}
\end{document}